\documentclass[useAMS,usenatbib,usegraphicx]{mn2e}
\usepackage{color}
\usepackage{psfig}
\usepackage{epsfig}
\usepackage{amsbsy}  
\usepackage{amssymb}

\definecolor{ndcol}{rgb}{0.75,0.0,0.0}

\usepackage[colorlinks,citecolor=black,linkcolor=black]{hyperref}

\newcommand*{\unit}[1]{\ensuremath{\mathrm{\, #1}}}
\newcommand*{\erg}{\unit{erg}}
\newcommand*{\keV}{\unit{keV}}
\newcommand*{\second}{\unit{s}}

\newcommand*{\cm}{\unit{cm}}

\newcommand*{\Mpc}{\unit{Mpc}}

\newcommand{\Hunit}{~{\rm km}~{\rm s}^{-1} {\rm Mpc}^{-1} }

\newcommand*{\E}[1]{\ensuremath{\times 10^{#1}}}
\newcommand*{\logTen}{\ensuremath{\log_{10}}}
\newcommand*{\expectation}[1]{\ensuremath{\left\langle #1 \right\rangle}}

\newcommand*{\Msun}{\ensuremath{\, M_{\odot}}}
\newcommand*{\mysub}[2]{\ensuremath{#1_{\mathrm{#2}}}}

\defcitealias{Rapetti:09}{R09}
\defcitealias{Mantz:08}{M08}
\defcitealias{Allen:08}{A08}
\newcommand*{\depaper}{Paper~I}
\newcommand*{\scpaper}{Paper~II}

\newcommand*{\ROSAT}{{\it{ROSAT}}}

\title[Testing General Relativity with the observed growth of massive clusters] 
{The Observed Growth of Massive Galaxy Clusters III: Testing General Relativity on Cosmological Scales}
\author[D.~Rapetti, S.~W.~Allen, A.~Mantz \& H.~Ebeling] {David~Rapetti${}^{1}$\thanks{Email: drapetti@slac.stanford.edu},
  Steven~W.~Allen${}^{1}$, Adam~Mantz${}^{1,2}$ and Harald~Ebeling$^3$\\
  ${}^1$ Kavli Institute for Particle Astrophysics and Cosmology at Stanford University, 452 Lomita Mall, Stanford 94305-4085, CA, USA \\
  and SLAC National Accelerator Laboratory, 2575 Sand Hill Road, Menlo Park 94025, CA, USA\\
  $^2$NASA Goddard Space Flight Center, Code 662, Greenbelt, MD 20771, USA\\
  $^3$Institute for Astronomy, 2680 Woodlawn Drive, Honolulu, HI 96822, USA}

\begin{document}
\date{Accepted ???, Received ???; in original form \today}
\pagerange{\pageref{firstpage}--\pageref{lastpage}} \pubyear{2008}

\maketitle
\label{firstpage}

\begin{abstract}
  This is the third of a series of papers in which we derive
  simultaneous constraints on cosmological parameters and X-ray
  scaling relations using observations of the growth of massive, X-ray
  flux-selected galaxy clusters. Our data set consists of 238 clusters
  drawn from the {\it ROSAT} All-Sky Survey, and incorporates
  extensive follow-up observations using the {\it Chandra} X-ray
  Observatory. Here we present improved constraints on departures from
  General Relativity (GR) on cosmological scales, using the growth
  index, $\gamma$, to parameterize the linear growth rate of cosmic
  structure. Using the method of \cite{Mantz:09a}, we simultaneously
  and self-consistently model the growth of X-ray luminous clusters
  and their observable-mass scaling relations, accounting for survey
  biases, parameter degeneracies and the impact of systematic
  uncertainties. Such analysis of the survey and follow-up data is
  crucial, else spurious constraints may be obtained. We combine the
  X-ray cluster growth data with cluster gas mass fraction, type Ia
  supernova, baryon acoustic oscillation and cosmic microwave
  background data. We find that the combination of these data leads to
  a tight correlation between $\gamma$ and the normalization of the
  matter power spectrum, $\sigma_{8}$. Consistency with GR requires a
  measured growth index of $\gamma\sim 0.55$. Under the assumption of
  self-similar evolution and constant scatter in the cluster
  observable-mass scaling relations, and for a spatially flat model
  with a cosmological constant, we measure
  $\gamma(\sigma_{8}/0.8)^{6.8}=0.55^{+0.13}_{-0.10}$, with allowed
  values for $\sigma_{8}$ in the range $0.79$ to $0.89$ (68.3 per cent
  confidence limits). Relaxing the assumptions on the scaling
  relations by introducing two additional parameters to model possible
  evolution in the normalization and scatter of the luminosity-mass
  relation, we obtain consistent constraints on $\gamma$ that are only
  $\sim 20$ per cent weaker than those above. Allowing the dark energy
  equation of state, $w$, to take any constant value, we
  simultaneously constrain the growth and expansion histories, and
  find no evidence for departures from either GR or the cosmological
  constant plus cold dark matter paradigm. Our results represent the
  most robust consistency test of General Relativity on cosmological
  scales to date.
\end{abstract}

\begin{keywords}
  cosmological parameters -- cosmology: observations -- cosmology:
  theory -- X-ray: galaxies: clusters
\end{keywords}

\section{Introduction}
\label{sec:introduction}

Since \cite{Riess:98} and \cite{Perlmutter:99} first demonstrated,
using observations of type Ia supernovae (SNIa), that the expansion
history of the Universe is accelerating at late times, improved SNIa
studies have continuously corroborated these results
\citep{Garnavich:98, Schmidt:98, Knop:03, Tonry:03, Barris:04,
  Riess:04, Riess:07, Astier:06, Wood-Vasey:07, Kowalski:08,
  Hicken:09}. At the same time, measurements of anisotropies in the
cosmic microwave background (CMB) with the Wilkinson Microwave
Anisotropy Probe \citep[WMAP][and companion
papers]{Spergel:03,Spergel:07,Komatsu:09, Dunkley:09} and other CMB
experiments \citep{Readhead:04, Jones:06, Reichardt:09, Chiang:09,
  Gupta:09} have tightly constrained other key cosmological
parameters. A variety of other cosmological data sets have also been
used to independently measure cosmic acceleration. Measurements of the
gas mass fraction ($f_{\rm gas}$) in galaxy clusters, having earlier
shown that the mean matter density of the Universe is low,
$\Omega_{\rm m}\sim 0.25$ \cite[see e.g.][]{White:93}, have directly
confirmed the effects of cosmic acceleration, at comparable
significance to that seen in SNIa data \citep[see e.g.][]{Allen:04,
  Allen:08}. Measurements of baryon acoustic oscillations (BAO) in
galaxy surveys \citep[see e.g.][]{Eisenstein:05, Percival:07,
  Percival:09}, the correlation of the low multipoles of the CMB with
large scale structure observations from various surveys \citep[see
e.g.][]{Scranton:03, Fosalba:03, Cabre:06, Giannantonio:08, Ho:08},
and weak gravitational lensing by large scale structure
\citep{Schrabback:09} have also shown evidence for cosmic
acceleration. The combination of subsets and/or all of these data have
led to the establishment of the $\Lambda$CDM paradigm, in which the
energy density of the Universe is currently composed of $\sim 5$ per
cent baryons, $\sim 20$ per cent cold dark matter (CDM) and $\sim 75$
per cent dark energy, the last of which drives cosmic acceleration and
has identical characteristics to Einstein's cosmological constant,
$\Lambda$.

Together, the above experiments robustly show that the Universe is
accelerating. However, the ability of such data to probe the
underlying cause of this acceleration is limited. All of the above
constraints on dark energy are primarily driven by its effects on the
background geometry. Such data alone cannot distinguish acceleration
due to a true dark energy component with negative pressure from
modifications to the standard theory of gravity, i.e. Einstein's
General Relativity (GR). Recently, however, \citet[hereafter
\depaper{}]{Mantz:08, Mantz:09a}, and \cite{Vikhlinin:09} have
presented new constraints on dark energy from measurements of the
growth of cosmic structure, as evidenced in X-ray flux-selected
cluster samples. These experiments are sensitive not only to the
evolution of the mean energy density background, but also to the
evolution of the density perturbations with respect to this
background. This work has provided clear, independent confirmation of
the effects of dark energy in slowing the growth of X-ray luminous
galaxy clusters.

The observed evolution of density perturbations provides a sensitive
probe of the underlying theory of gravity and the
clustering\footnote{Note that for models other than the cosmological
  constant, dark energy is expected to couple with gravity,
  i.e. cluster \cite[see e.g.][]{Hu:05, Mota:07}. For specific dark
  energy and modified gravity models, see the reviews of
  \cite{Copeland:06} and \cite{Frieman:08}.} properties of dark
energy. \citet[hereafter \citetalias{Rapetti:09}]{Rapetti:09}
exploited this sensitivity to constrain departures from the predicted
cosmic growth rate for GR using the convenient parameterization
$\Omega_{\rm m}(z)^{\gamma}$, for which GR has $\gamma\sim 0.55$
\citep{Peebles:80, Wang:98, Linder:05, Huterer:06, DiPorto:07,
  Sapone:07, Gannouji:08, Wei:08,Linder:07, Nesseris:07}. Using the
growth of structure analysis of \cite{Mantz:08},
\citetalias{Rapetti:09} found no evidence for deviations from either
GR or $\Lambda$CDM. Recent results from the combination of other
cosmological data sets are also consistent with GR \citep{Daniel:10,
  Zhao:10, Reyes:10}.

This is the third of a series of papers in which, for the first time,
we employ a fully self-consistent analysis of the growth of massive
clusters that combines current X-ray cluster surveys and deep,
pointed, follow-up observations to simultaneously constrain both
cosmological parameters and observable-mass (luminosity-mass and
temperature-mass) scaling relations. Importantly, the improved method,
which is described in \depaper{}, properly accounts for all selection
biases, covariances and parameter degeneracies, and models fully the
impact of systematic uncertainties. Here, we utilize this improved
method (hereafter XLF, as defined in \depaper{}) to re-investigate the
simultaneous constraints that can be placed on $\gamma$ and the
background evolution (expansion history). For the background, we use
two reference expansion models: flat $\Lambda$CDM, parameterized by
the mean matter density, $\Omega_{\rm m}$; and flat $w$CDM,
parameterized by $\Omega_{\rm m}$ and a constant dark energy equation
of state, $w$. Our constraints on $\gamma$ arise primarily from the
XLF experiment, which uses the cluster samples of \citet[][\ROSAT{}
Brightest Cluster Sample; BCS]{Ebeling:98}, \citet[][\ROSAT{}-ESO Flux
Limited X-ray sample; REFLEX]{Bohringer:04}, \citet[][2009 in
preparation, Bright MAssive Cluster Survey; Bright MACS]{Ebeling:01},
plus extensive X-ray follow-up data from the {\it Chandra} X-ray
Observatory and {\it ROSAT} \cite[for details see][hereafter
\scpaper]{Mantz:09b}, although we also utilize the (currently) small
additional constraining power available from measurements of the
Integrated Sachs-Wolfe (ISW) effect at low multipoles in the CMB.

We emphasize that the models used for the evolution of the background
and density perturbations are purely phenomenological, encompassing
both modified gravity and clustering dark energy as possible sources
for cosmic acceleration. Our analysis allows for a simple, but elegant
test for departures from the standard GR and $\Lambda$CDM
paradigms. Using the combination of XLF, $f_{\rm gas}$, SNIa, BAO and
CMB, we demonstrate good agreement with the standard GR+$\Lambda$CDM
model and provide tight constraints on the model parameters. We show
that our results are robust against reasonable assumptions regarding
the evolution of galaxy cluster scaling relations. We stress that in
order to obtain robust results from current and future XLF studies, it
is essential to employ a consistent analysis method, such as the one
used here (see details in \depaper{}), in order to properly account
for selection biases, covariances, parameter degeneracies and
systematic uncertainties. Otherwise, spuriously tight constraints may
be obtained.

\section{Cosmological model}
\label{sec:cosmo}

\subsection{Linear growth rate}
\label{sec:gi}

Galaxy clusters are the largest and rarest virialized objects in the
Universe. Their abundance as a function of redshift provides an
extremely sensitive probe of the underlying cosmology. To predict the
cumulative number density of galaxy clusters, $n(M,z)$, at a given
mass, $M$, and redshift, $z$, we calculate the evolution of density
fluctuations using linear perturbation theory, accounting for
non-linear effects using results from N-body simulations.

The cumulative mass function of dark matter halos
can be calculated as

\begin{equation}
  n(M,z)=\int_{0}^{M}f(\sigma)\,\frac{\bar{\rho}_{\rm m}}{M'}\,\frac{d\ln\sigma^{-1}}{dM'}\,dM'\,,
\label{eq:nmz}
\end{equation}

\noindent where $\bar{\rho}_{\rm m}$ is the mean comoving matter
density, $\sigma^2$ is the variance of the linearly evolved density
field (as defined in Equation~\ref{eq:var}), and $f(\sigma)$ is a
fitting formula obtained from theory or N-body simulations
\citep{Press:74,Bond:91,Sheth:99,Jenkins:01, Evrard:02}. Remarkably,
$f(\sigma)$ encompasses linear and non-linear effects in such a way
that its form is approximately independent of the cosmology assumed
(see Section~\ref{sec:nonlinear}).

Modifications to GR at large scales have been proposed as alternatives
to dark energy to explain cosmic acceleration at late times. In
general, in addition to changing the expansion history, such
modifications also affect the growth of cosmic structure. As in
\citetalias{Rapetti:09}, we parameterize the growth rate, $f(a)$,
using \citep{Peebles:80, Wang:98, Linder:07}

\begin{equation}
  f(a)\equiv \frac{d\ln\delta}{d\ln a}=\Omega_{\rm m} (a)^{\rm
    \gamma}\,,
\label{eq:fa}
\end{equation}

\noindent where $\Omega_{\rm m}(a)=\Omega_{\rm m}
a^{-3}/E(a)^{2}$. Here, $\Omega_{\rm m}$ is the present, mean matter
density in units of the critical density, $E(a)=H(a)/H_0$ the
evolution parameter, $H(a)$ the Hubble parameter (with $H_0$ its
present-day value), and $\gamma$ the growth index.

We calculate the evolution of the linear matter density contrast,
$\delta\equiv \delta\rho_{\rm m}/\bar{\rho}_{\rm m}$, i.e. the ratio
of matter density fluctuations to the cosmic mean, by solving
Equation~\ref{eq:fa}. As an initial condition for this differential
equation, we match the value of $\delta$ at early times ($z_{\rm
  t}=30$) to that of GR. Given $\delta(z)$, we obtain the growth
factor between $z_{\rm t}$ and $z$,

\begin{equation}
  D(z)\equiv\frac{\delta(z)}{\delta(z_{\rm t})}\,.
\label{eq:growthf}
\end{equation}

\noindent Note that $z_{\rm t}=30$ is well within the dark matter
dominated era, when the expansion is expected to be decelerating. At
this time $f(a)$ is close to 1, independent of $\gamma$. Using
$D(z)$, we calculate the linear matter power spectrum for a given
wavenumber, $k$, and redshift,

\begin{equation}
  P(k,z)\propto k^{\rm n_{\rm s}} T^2(k,z_{\rm t}) D(z)^2\,,
\label{eq:powers}
\end{equation}

\noindent where $T(k,z_{\rm t})$ is the matter transfer function of GR
at redshift $z_{\rm t}$, and $n_{\rm s}$ the primordial scalar
spectral index. Given $P(k,z)$, the variance of the linearly evolved
density field, smoothed by a spherical top-hat window function of
comoving radius $R$ enclosing mass $M=4\pi\bar{\rho}_{\rm m}R^3/3$, is
then
 
\begin{equation}
  \sigma^2(M,z) = \frac{1}{2\pi^2} \int_0^\infty k^2 P(k,z) |W_{\rm M}(k)|^2 dk\,,
\label{eq:var}
\end{equation}

\noindent where $W_{\rm M}(k)$ is the Fourier transform of the window
function. Using this value for $\sigma(M,z)$ in Equation~\ref{eq:nmz},
we obtain $n(M,z)$.

\subsection{Background evolution}
\label{sec:exp}

Equation~\ref{eq:fa} depends on the evolution of the background energy
density, through the parameter $E(a)$. Assuming an expansion history
appropriate for $\Lambda$CDM, it has been shown \citep[see
e.g.][]{Linder:07, Nesseris:07} that, for $\gamma\sim0.55$, the growth
rate calculated using Equation~\ref{eq:fa} corresponds remarkably
accurately to that predicted by GR. \cite{Linder:07} also showed that
$\gamma$ and $w$ are only weakly correlated (see also
\citetalias{Rapetti:09}) and, therefore, that one can measure the
cosmic growth and expansion rates almost independently. These
convenient properties of the $\gamma$-model provide us with a simple
but powerful consistency test for GR.

Here, we perform this test using the new cluster XLF data (Papers I
and II) in combination with $f_{\rm gas}$, SNIa, BAO and CMB
constraints. The latter four data sets primarily constrain the
background evolution of the Universe, i.e. the expansion history. As a
background expansion model, we use\footnote{\citetalias{Rapetti:09}
  examined three expansion models: flat $\Lambda$CDM, flat $w$CDM and
  non-flat $\Lambda$CDM. For the latter, they found that there is
  negligible covariance between $\Omega_{\rm k}$, the curvature energy
  density, and $\gamma$. Since this test is computationally expensive,
  we do not repeat it here.}

\begin{equation}
  E(a)=\left[\Omega_{\rm m} \,a^{-3}+(1-\Omega_{\rm m}) \,a^{-3(1+w)}\right]^{1/2}\,,
\label{eq:Ea}
\end{equation}

\noindent where $w$ is the dark energy equation of state parameter and
$w=-1$ corresponds to flat $\Lambda$CDM. Note that we use this model {\it
  only} as a phenomenological description of the expansion history. We
do {\it not} assume the existence of dark energy. Our model allows us
to simultaneously test for departures from $\Lambda$CDM in the
expansion history ($w\neq-1$) and departures from GR in the growth
history ($\gamma\neq0.55$). Deviations from these benchmarks can be
caused either by clustering dark energy or modified gravity.

\subsection{Non-linear model}
\label{sec:nonlinear}

The fitting formula $f(\sigma)$ from \cite{Jenkins:01} has been tested
for a wide range of background dark energy cosmologies and has been
shown to provide a ``universal'' description, within $\sim 20$ per
cent uncertainty \cite[see e.g.][]{Kuhlen:04, Lokas:04, Klypin:03,
  Linder:03b, Mainini:03, Maccio:04, Francis:09,
  Grossi:09}\footnote{\cite{Schmidt:09a} explored an $f(R)$ gravity
  model and showed that for values of their model parameter compatible
  with Solar Systems tests ($f_{\rm R0}\sim10^{-6}$), the halo mass
  function is in agreement with that of GR for very massive clusters
  (within the quoted level of uncertainty of the
  simulations). However, for more extreme values of the model
  parameter and for less massive clusters (for which chameleon effects
  are important), significant differences can arise. For the
  self-accelerating branch of the braneworld gravity model DGP
  \citep{Dvali:00}, \cite{Schmidt:09} and \cite{Chan:09} predict a
  significant suppression of the halo mass function relative to GR,
  especially for massive objects. \cite{Schmidt:09b} predicts an
  enhancement for the normal branch of this model.}. However, in these
tests, typically only the background evolution was modified, while the
linear density perturbations of the new component (dark energy) were
neglected. Recently, \cite{Jennings:09} have presented results that
account for the effects of the density fluctuations provided by
quintessence models on the linear matter power spectrum, finding only
a small effect on $f(\sigma)$ for the masses and redshifts relevant
here.

The present study aims to test the consistency of current observations
with GR (and $\Lambda$CDM). Therefore, we use an $f(\sigma)$
corresponding to GR, seeking to determine whether values of
$\gamma\sim0.55$ (and $w=-1$) are preferred by the data. Beyond the
mass function of \cite{Jenkins:01}, several authors have investigated
the inclusion of an explicit dependence on redshift in the fitting
formula $f(\sigma,z)$ \citep{Lukic:07, Reed:07, Cohn:08}. Following
\depaper{}, we use the $f(\sigma,z)$ function of \cite{Tinker:08},
determined from a large suite of simulations,
 
\begin{equation}
  f(\sigma,z) = A \left[\left(\frac{\sigma}{b}\right)^{-a}+1\right]e^{-c/\sigma^2}\,.
\label{eq:mf}
\end{equation}

\noindent Here, parameters have the redshift dependence
$x(z)=x_0(1+z)^{\varepsilon \alpha_{\rm x}}$, with $x \in
\{A,a,b,c\}$. The values for $x_0$ and $\alpha_{\rm x}$ are given by
\cite{Tinker:08}. As described in \depaper{}, we use the parameter
$\varepsilon$ to account for residual systematic uncertainties in the
evolution of $f(\sigma,z)$. We marginalize over the uncertainties in
the parameters of Equation~\ref{eq:mf}, accounting for their
covariance. We also account for additional systematic uncertainties
due to e.g. the presence of baryons (for details, see
\depaper{}). Note, however, that, as we show in \depaper{}, the
statistical and systematic uncertainties in $f(\sigma,z)$ are
subdominant in the analysis. We have verified that $\varepsilon$ is
essentially uncorrelated with $\gamma$.

\section{Evolution of the scaling relations}
\label{sec:scal}

To compare the XLF data with the theoretical predictions of the mass
function, we need to relate mass, $M$, to our observables: the X-ray
luminosity, $L$, and average temperature, $T$, of the clusters. In
\scpaper{}, we describe the follow-up X-ray observations used to
determine these relations. As discussed in \depaper{}, we
simultaneously and self-consistently constrain both cosmological (see
Section~\ref{sec:cosmo}) and scaling relation parameters.

Following the notation of \scpaper{}, we model the evolution of the
luminosity-mass scaling relation as

\begin{equation}
  \label{eq:ml}
  \expectation{\ell(m)} = \beta_0^{\ell m} + \beta_1^{\ell m} m + \beta_2^{\ell m}\logTen(1+z)\,,
\end{equation}

\noindent with a log-normal, and possibly evolving, intrinsic scatter
of the luminosity at a given mass~\footnote{In the notation of
  \citetalias{Rapetti:09}, $\beta^{\ell m}_{0}$, $\beta^{\ell m}_{1}$,
  $\beta^{\ell m}_{2}$, $\sigma_{\ell m}$ and $\sigma_{\ell m}'$
  correspond to $\logTen(M_{0})$, $\beta$, $\zeta$, $\eta_0$ and
  $\eta_z$, respectively.}

\begin{equation}
  \label{eq:scat}
  \sigma_{\ell m}(z) = \sigma_{\ell m} ( 1 + \sigma_{\ell m}' z )\,.
\end{equation}

\noindent Here, $\ell
\equiv\logTen[L_{500}E(z)^{-1}/10^{44}\erg\second^{-1}]$ and
$m\equiv\logTen[E(z)M_{500}/10^{15}\Msun]$, where the subscript $500$
refers to quantities measured within radius $r_{500}$, at which the
mean, enclosed density is 500 times the critical density of the
Universe at redshift $z$. The factors of $E(z)$ are required to
account for the background evolution of the critical density. Fixing
$\beta_2^{\ell m}=0$, one has ``self-similar'' evolution of the
scaling relation \citep{Kaiser:86,Bryan:98}, determined entirely by
the $E(z)$ factors. Fixing $\sigma_{\ell m}'=0$, one has constant
scatter. In \scpaper{}, we show that departures from self-similar
evolution and evolution in $\sigma_{\ell m}(z)$ are not required by
current data.

In this paper we present results both for pure self-similarity and
constant scatter ($\beta_2^{\ell m}=0$, $\sigma_{\ell m}'=0$), and
also allowing for additional evolution in the luminosity-mass relation
and its scatter, through $\beta_2^{\ell m}$ and $\sigma_{\ell m}'$. As
we shall show, our improved analysis method (of \depaper{}) and the
inclusion of high quality follow-up data for a significant fraction of
clusters allows us to address even general questions of this type. We
emphasize that for such analysis it is essential to {\it
  simultaneously} model the mass function, scaling relations, growth
history and the impact of systematic uncertainties fully, else
spurious constraints may be obtained.

The data and analysis of \depaper{} and II also include measured
temperatures to simultaneously constrain the temperature-mass
relation. In \scpaper{}, we show that a simple power law

\begin{equation}
  \label{eq:mt}
  \expectation{t(m)} = \beta_0^{tm} + \beta_1^{tm} m\,,
\end{equation}

\noindent where $t\equiv \logTen\left(\mysub{kT}{500}/\keV\right)$,
without additional evolution parameters, such as $\beta_2^{tm}$ and
$\sigma_{tm}'$ (which are defined equivalently to the corresponding
parameters in the luminosity-mass relation, $\beta_2^{\ell m}$ and
$\sigma_{\ell m}'$), is sufficient to describe the data. Note also
that, for clusters with $kT>3\keV$, such as those used here, the flux
within the $0.1-2.4\keV{}$ band is nearly independent of the
temperature, and therefore $\beta_2^{tm}$ and $\sigma_{tm}'$ are
essentially uncorrelated with $\gamma$.\footnote{We have explicitly
  verified this to be the case.} Thus, we do not further consider
these parameters, and assume that the temperature-mass relation
evolves self-similarly, with a constant log-normal
scatter\footnote{Our analysis includes a correlation parameter between
  the intrinsic scatters of the luminosity-mass and temperature-mass
  relations, $\rho_{\ell tm}$ (see details in \scpaper{}).}.

\begin{figure*}
\begin{center}
\includegraphics[width=3.2in]{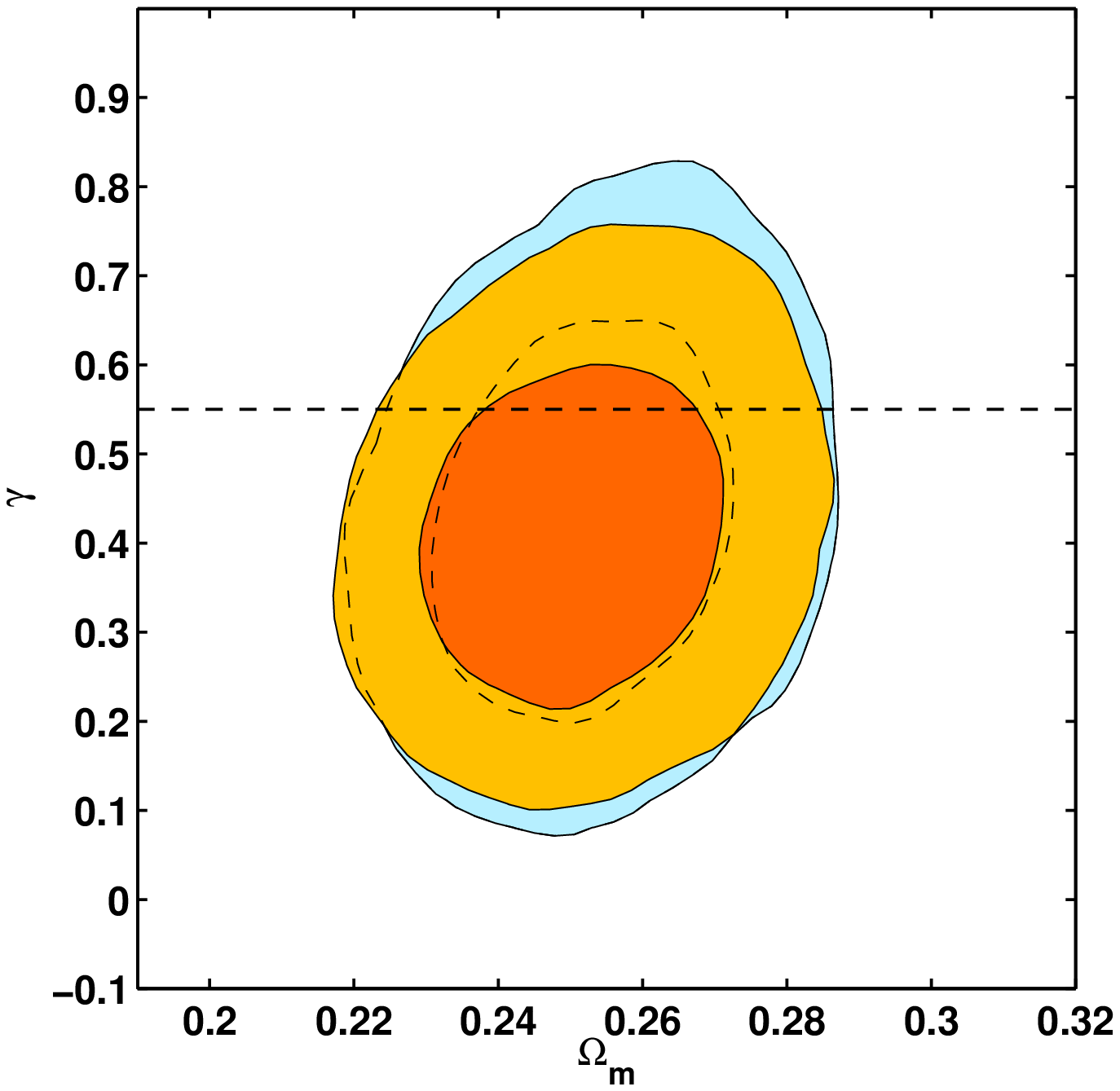}
\hspace{0.6cm}
\includegraphics[width=3.2in]{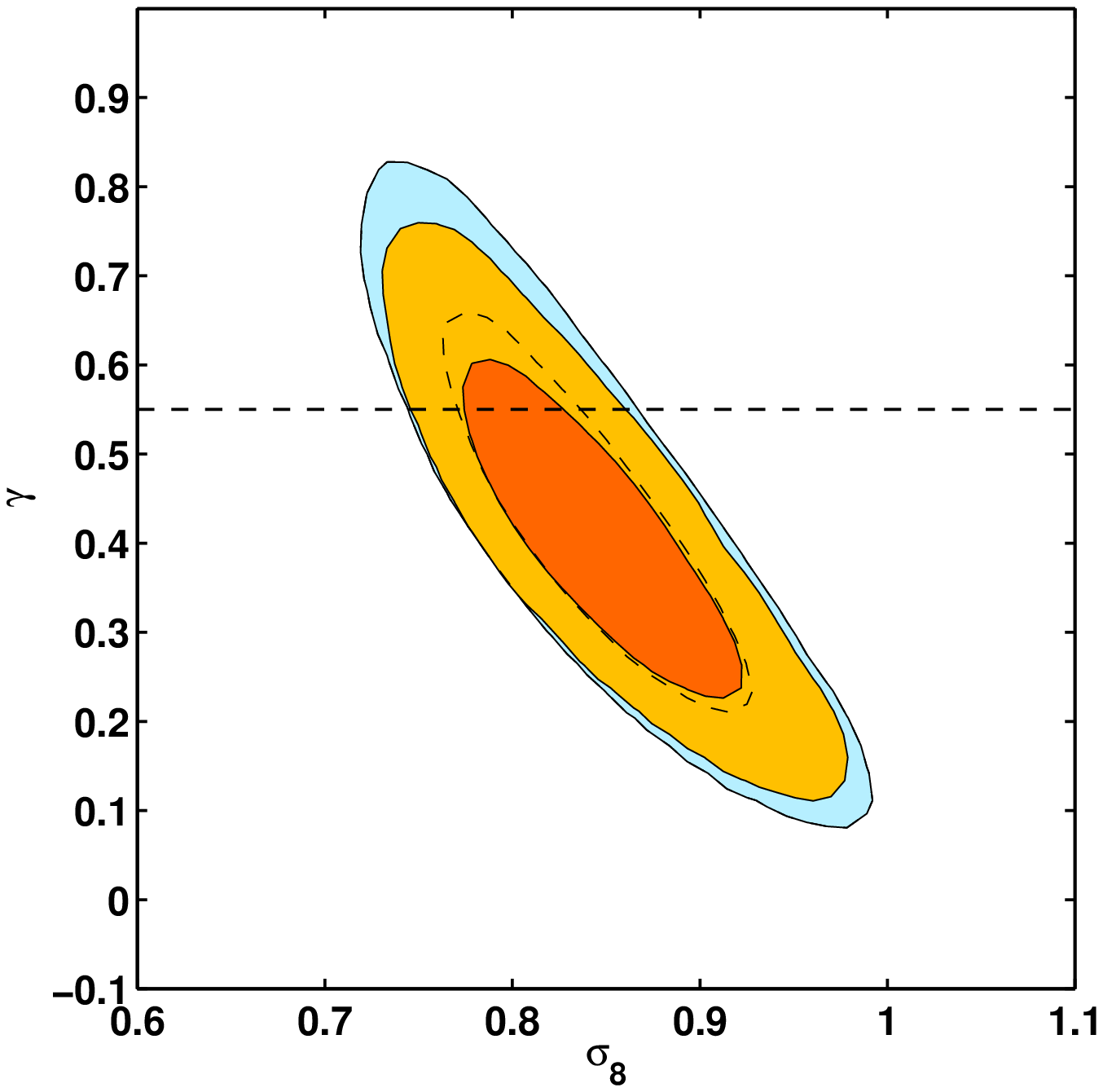}
\caption{68.3 and 95.4 per cent confidence contours in the
  $\Omega_{\rm m},\gamma$ (left panel) and $\sigma_8, \gamma$ (right
  panel) planes for an assumed flat $\Lambda$CDM background model,
  using the combination of XLF, $f_{\rm gas}$, SNIa, BAO and CMB
  data. The gold, smaller contours assume self-similar evolution of
  the observable-mass scaling relations and constant scatter
  ($\beta_2^{\ell m}=0$; $\sigma_{\ell m}'=0$). The blue, larger
  contours allow for departures from self-similarity and redshift
  evolution in the scatter of the luminosity-mass relation
  (Section~\ref{sec:scal}; \scpaper{}). The horizontal, dashed lines
  mark $\gamma=0.55$, the growth index for GR.}
\label{fig:lcdm}
\end{center}
\end{figure*}

\section{ISW effect}
\label{sec:isw}

Our analysis also accounts for the small, but non-negligible
additional constraining power on $\gamma$ that is available from
measurements from the ISW effect. The low multipoles of the CMB are
sensitive to the growth of structure through this effect. The time-varying
gravitational potentials of large scale structures contribute a net
effect on the energy of the photons crossing them. For these photons,
we calculate their contribution to the temperature anisotropy power
spectrum as \citep{Weller:03}

\begin{equation}
  \Delta_{l}^{\rm ISW}(k) = 2 \int dt\, {\rm e}^{-\tau(t)}
  \phi'j_{l}\left[k(t-t_{\rm 0})\right]\,,
  \label{eq:deltaisw}
\end{equation}

\noindent where $t$ is the conformal time, $\tau$ the optical depth to
reionization, $j_{l}(x)$ the spherical Bessel function for the
multipole $l$, and $\phi'$ the conformal time variation of the
gravitational potential. The latter can be calculated in terms of
$\gamma$ using the Poisson equation, $k^2\phi=-4 \pi G
a^2\,\delta\rho_{\rm m}$, as (\citetalias{Rapetti:09})

\begin{equation}
  \phi' = \frac{4 \pi G a^2}{k^2}\,\mathcal{H}\,\delta\bar{\rho}_{\rm m}\left[1-\Omega_{\rm m}(a)^{\gamma}\right]\,,
\label{eq:isw}
\end{equation}

\noindent where $\mathcal{H}$ is the conformal Hubble parameter. Since
we are performing a consistency test of GR, we assume that, as in GR,
the contributions of the anisotropic stress and energy flux are
negligible \cite[for details on these terms
see][]{Challinor:98}\footnote{For particular alternative gravity
  models such as DGP, these terms will not be negligible \citep{Hu:07}
  and the Poisson equation will need to be modified
  \citep{Hu:08}.}. As an initial condition, at $z=2$ we match the
$\Delta_{l}^{\rm ISW}(k)$ to that of GR, using Equation~\ref{eq:isw}
to evolve the model to $z=0$. Unless stated, for results including CMB
data, we include the constraints on $\gamma$ available from the ISW
effect.

\begin{figure*}
\begin{center}
\includegraphics[width=4.5in]{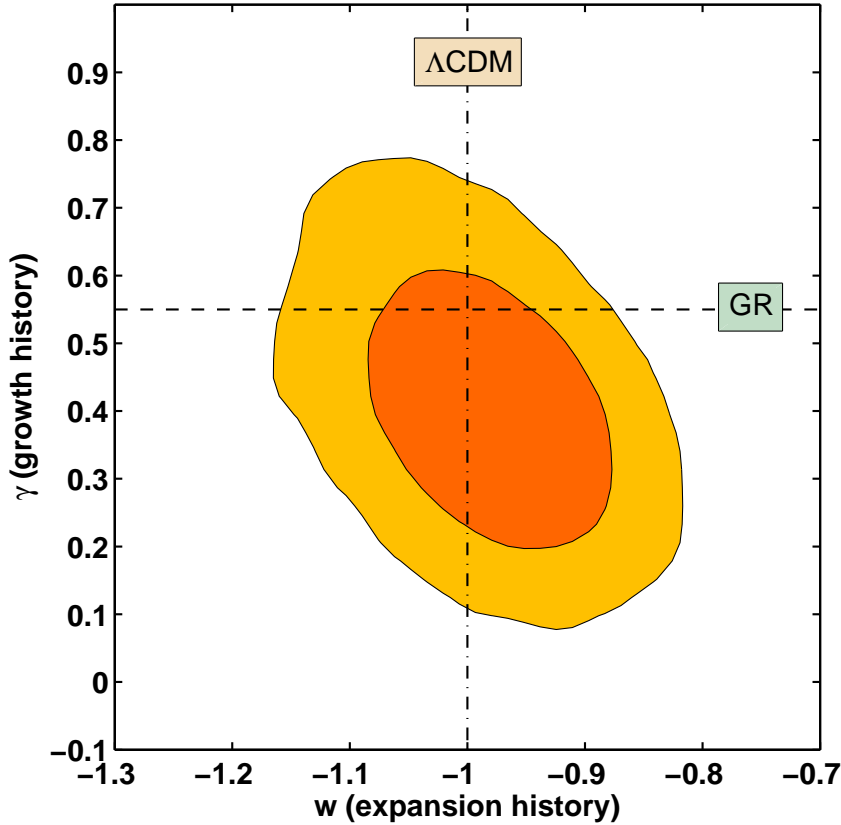}
\caption{68.3 and 95.4 per cent confidence contours in the $w,\gamma$
  plane for the flat $w$CDM background expansion model, using the
  combination of XLF, $f_{\rm gas}$, SNIa, BAO and CMB data. The
  results assume self-similar evolution of the scaling relations and
  constant scatter ($\beta_2^{\ell m}=0$; $\sigma_{\ell m}'=0$). The
  horizontal, dashed line marks $\gamma=0.55$, the growth history for
  GR. The vertical, dotted-dashed line marks $w=-1$, the expansion
  history for $\Lambda$CDM. The results are simultaneously consistent
  with GR and $\Lambda$CDM.}
\label{fig:wconst}
\end{center}
\end{figure*}

\section{Data analysis}
\label{sec:data}

To constrain the cosmic growth rate we use three wide-area cluster
samples drawn from the \ROSAT{} all-sky survey: BCS ($z<0.3$; northern
sky; $F_{\rm X}(0.1-2.4\keV{})>4.4\times 10^{-12} \erg \second^{-1}
\cm^{-2}$), REFLEX ($z<0.3$; southern sky; $F_{\rm X}>3.0\times
10^{-12} \erg\second^{-1} \cm^{-2}$), and Bright MACS ($0.3<z<0.5$;
$\sim 55$ per cent sky coverage; $F_{\rm X}>2\times 10^{-12} \erg
\second^{-1} \cm^{-2}$). In order to keep systematic uncertainties to a
minimum, for all three surveys, we impose a lower luminosity cut of
$2.5\E{44} h_{70}^{-2}\erg\second^{-1}$ ($0.1-2.4\keV{}$). For the
BCS, this gives a total of 78 clusters; for REFLEX, 126 clusters; and
for Bright MACS, 34 clusters. In total we have 238 clusters.

To constrain the background evolution we also use measurements of the
gas mass fraction ($f_{\rm gas}$) in 42 massive, dynamically relaxed
clusters from \cite{Allen:08}, as well as the compilation of 307 SNIa
from \cite{Kowalski:08}, and BAO data \citep{Percival:07, Colless:01,
  Colless:03, Adelman:07}. The latter constrain the ratio of the sound
horizon to the distance scale at $z=0.25$ and $z=0.35$. We also use
the five-year WMAP data \cite[][and companion papers]{Dunkley:09,
  Komatsu:09}.

We calculate the posterior probability distributions of all parameters
using the Metropolis Markov Chain Monte Carlo (MCMC) algorithm, as
implemented in the {\sc
  CosmoMC}\footnote{http://cosmologist.info/cosmomc/} code of
\cite{Lewis:02}. We use a modified version of this code that includes
additional modules to calculate the likelihood for our two cluster
experiments: $f_{\rm
  gas}$\footnote{http://www.stanford.edu/$\sim$drapetti/fgas\_module/}
\citep{Rapetti:05,Allen:08} and the XLF (\depaper{}). For analyses
without CMB data, we use standard Gaussian priors on both the
present--day Hubble constant, $H_0=72\pm8\Hunit$ \citep{Freedman:01},
and the mean physical baryon density, $\Omega_{\rm
  b}h^2=0.0214\pm0.0020$ \citep{Kirkman:03}, from Big Bang
Nucleosynthesis studies. When including the CMB data, we do not use
these priors.

When using CMB data, instead of $H_0$ we fit $\theta$. The latter is
the (approximate) ratio of the sound horizon at last scattering to the
angular diameter distance, which, as shown by \cite{Kosowsky:02}, is
less correlated than $H_0$ with other parameters. In all analyses, in
addition to $H_0$ or $\theta$, and $\Omega_{\rm b}h^2$, we fit for the
mean physical dark matter density, $\Omega_{\rm c}h^2$; the logarithm
of the adiabatic scalar amplitude, $\ln(A_{\rm s})$ (which is related
to $\sigma_{8}$)\footnote{$\sigma_8^2$ is the $z=0$ variance in the
  density field at scales of $8h^{-1}\Mpc$ (see
  equation~\ref{eq:var}).}; the growth index, $\gamma$; and, when
stated, the kinematical $w$ parameter\footnote{Recall that we use $w$
  purely as a kinematic parameter, allowing us to model the cosmic
  expansion history conveniently.}. For the analyses with CMB data, we
also fit the optical depth to reionization, $\tau$, the adiabatic
scalar spectral index, $n_{\rm s}$, and marginalize over the amplitude
of the Sunyaev-Zel'dovich effect from galaxy clusters, $0<A_{\rm SZ}<
2$ \citep{Spergel:07}. For analyses without CMB data, we fix $n_{\rm
  s}$ to $0.95$ since, for such analyses, $n_{\rm s}$ is degenerate
with $\sigma_{8}$ (see \depaper{}).

\section{Simultaneous constraints on the growth and expansion
  histories}
\label{sec:results}

\subsection{Results for the $\Lambda$CDM expansion model}
\label{sec:growth}

Figure~\ref{fig:lcdm} shows the joint constraints in the $\Omega_{\rm
  m}, \gamma$ (left panel) and $\sigma_8, \gamma$ (right panel) planes
for a flat $\Lambda$CDM expansion history. The gold (smaller) contours
show the constraints for the minimal, self-similar and
constant-scatter scaling relation model ($\beta^{lm}_2=0$ and
$\sigma_{\ell m}'=0$).

Figure~\ref{fig:lcdm} shows that, for the $\Lambda$CDM background
model, $\Omega_{\rm m}$ is not strongly correlated with $\gamma$. This
supports the idea that $\gamma$ is a useful parameterization to
measure departures from the growth rate of GR, approximately
independent of assumptions regarding the background evolution,
parameterized in this case by $\Omega_{\rm m}$.

The right panel of Figure~\ref{fig:lcdm} shows a clear correlation
between $\sigma_8$ and $\gamma$, with a correlation coefficient of
$\rho=-0.87$. From the MCMC samples we measure
$\gamma(\sigma_{8}/0.8)^{6.8}=0.55^{+0.13}_{-0.10}$, with allowed
values for $\sigma_{8}$ in the range $0.79<\sigma_{8}<0.89$
(marginalized 68.3 per cent confidence limits)\footnote{We calculate the
  exponent $6.8$ finding the best constrained eigenvector for the
  estimated covariance matrix between $\ln\gamma$ and
  $\ln\sigma_{8}$.}. For $\sigma_{8}=0.8$, we measure $\gamma=0.55$
(GR; marked by a horizontal, dashed line in
Figure~\ref{fig:lcdm}). Our results are consistent with GR at the 68.3
per cent confidence level. The observed correlation between
$\sigma_{8}$ and $\gamma$ shows that the addition of precise,
independent knowledge of $\sigma_8$ should significantly improve the
precision of the constraints on $\gamma$ (see also
Section~\ref{sec:datasets}).

\begin{figure*}
\begin{center}
\includegraphics[width=3.2in]{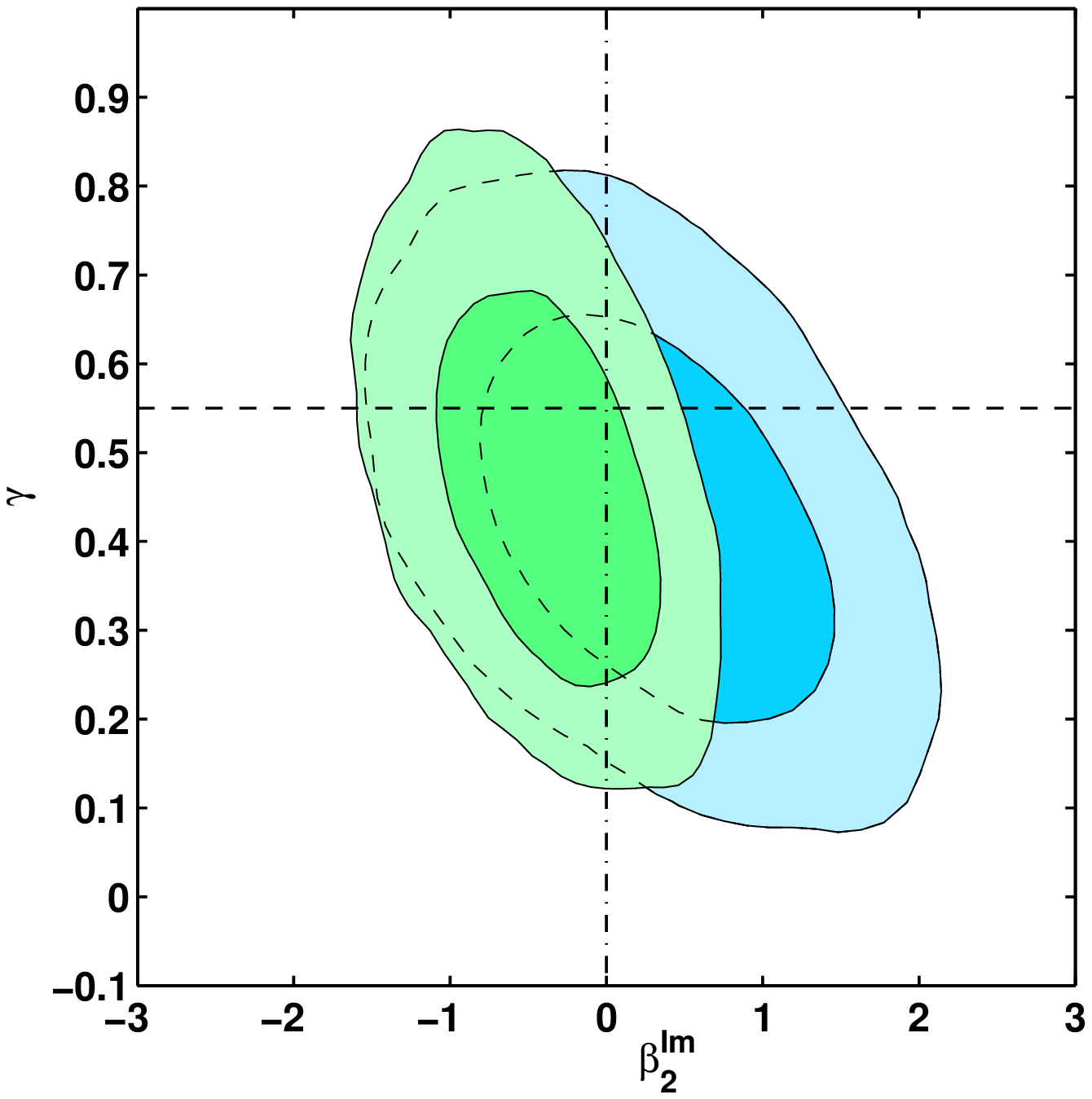}
\hspace{0.6cm}
\includegraphics[width=3.2in]{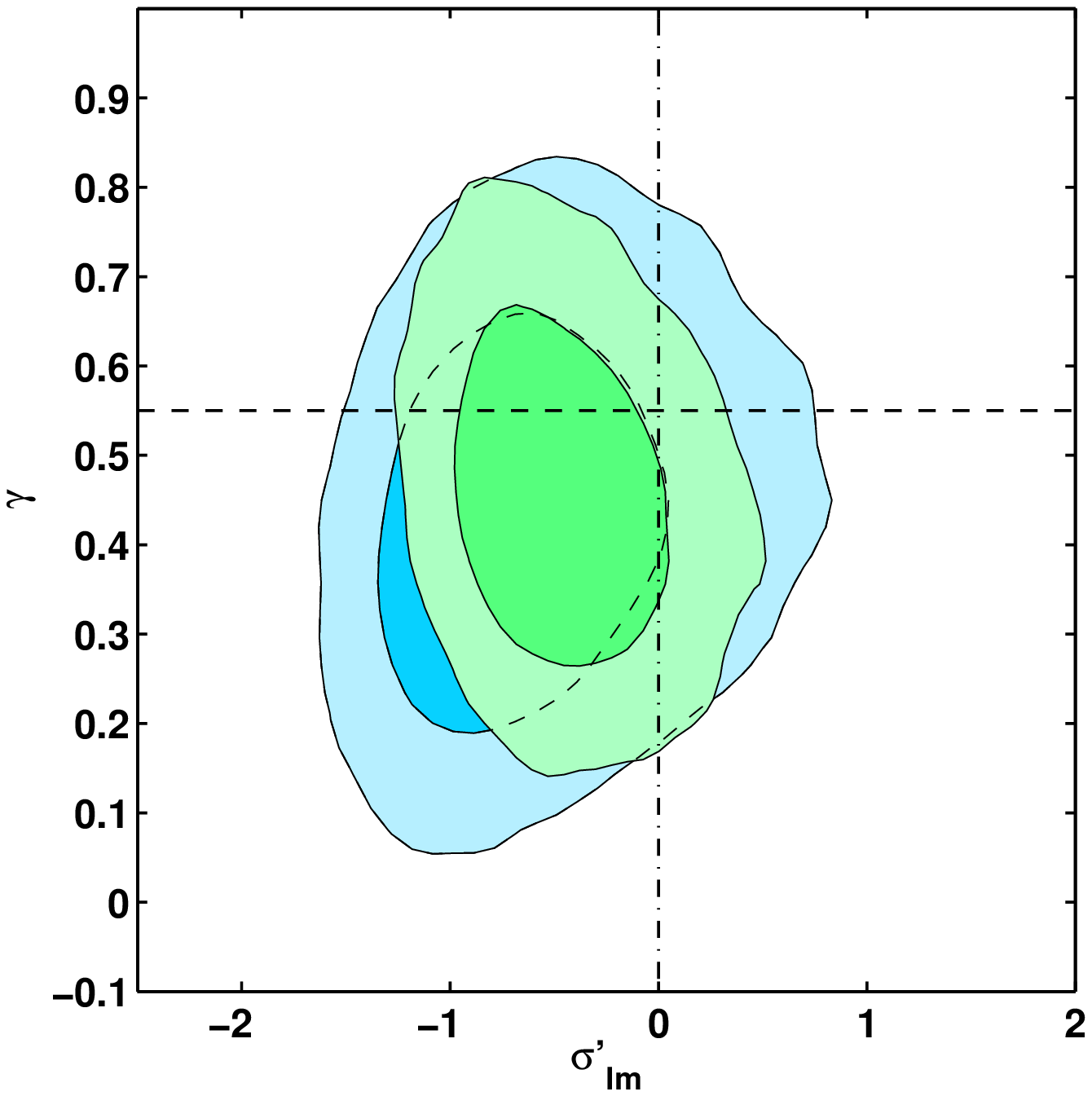}
\caption{68.3 and 95.4 per cent confidence contours in the
  $\beta^{\ell m}_2,\gamma$ (left panel) and $\sigma'_{\ell m},
  \gamma$ (right panel) planes for the flat $\Lambda$CDM background
  evolution model, from the combination of XLF, $f_{\rm gas}$, SNIa,
  BAO and CMB data. The green, smaller contours show the constraints
  obtained with either $\beta^{\ell m}_2$ (left panel) or
  $\sigma'_{\ell m}$ (right panel) allowed to be a free parameter. The
  blue, larger contours show the results when both parameters,
  $\beta^{\ell m}_2$ and $\sigma'_{\ell m}$, are allowed to be
  free. The horizontal, dashed lines mark $\gamma=0.55$ (GR). The
  vertical, dotted-dashed lines mark the self-similar ($\beta^{\ell
    m}_2=0$; left panel) and constant scatter ($\sigma'_{\ell m}=0$;
  right panel) conditions.}
\label{fig:scal}
\end{center}
\end{figure*}

The marginalized constraints on a single interesting parameter are
summarized in Table~\ref{table:params}.

\subsection{Results for the $w$CDM expansion model}
\label{sec:wcdm}

Figure~\ref{fig:wconst} shows the joint constraints in the $w, \gamma$
plane for the $w$CDM background cosmology ($w$ constant), assuming
self-similar evolution and constant scatter ($\beta_2^{\ell m}=0$;
$\sigma_{\ell m}'=0$). The horizontal, dashed and vertical,
dotted-dashed lines indicate $\gamma=0.55$ (GR) and $w=-1$
($\Lambda$CDM), respectively. Importantly, our results are
simultaneously consistent with both GR and $\Lambda$CDM at the 68.3
per cent confidence level (see Figure~\ref{fig:wconst}). For this
expansion model, we again observe a correlation between $\sigma_8$ and
$\gamma$, although somewhat less ($\rho=-0.69$) than for the
$\Lambda$CDM case. The marginalized constraints on a single
interesting parameters are summarized in Table~\ref{table:params}.

\begin{figure}
\begin{center}
\includegraphics[width=3.2in]{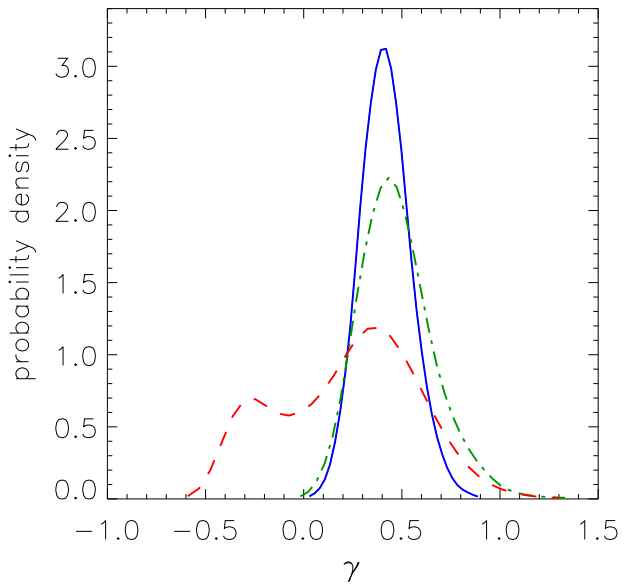}
\caption{Marginalized constraints on $\gamma$ for the flat
  $\Lambda$CDM background expansion model, from the XLF data alone
  (green, dotted-dashed line), from the combination of SNIa, $f_{\rm
    gas}$, BAO and CMB data (red, dashed line), and from all the data
  sets combined, including the XLF (blue, solid line). It is clear
  that the XLF dominates the constraints on $\gamma$ and that, alone,
  it provides significantly tighter constraints than those from the
  combination of the other data sets.}
\label{fig:comparison}
\end{center}
\end{figure}

\subsection{Testing for departures from self-similarity}
\label{sec:followup}

The blue (larger) contours in Figure~\ref{fig:lcdm} show the
constraints obtained for the $\Lambda$CDM background model when
allowing $\beta_2^{\ell m}$ and $\sigma_{\ell m}'$ to be
free. Remarkably, using this (significantly) more complex model for
the luminosity-mass relation, we obtain constraints on $\gamma$ that
are only $\sim20$ per cent weaker than those for the minimal,
self-similar, constant-scatter model (gold, smaller contours). The
primary reason for this robustness is the comprehensive nature of the
follow-up data in the XLF experiment, which tightly constrain the
scaling relations and their evolution (\scpaper{}).

\begin{figure*}
\begin{center}
\includegraphics[width=3.36in]{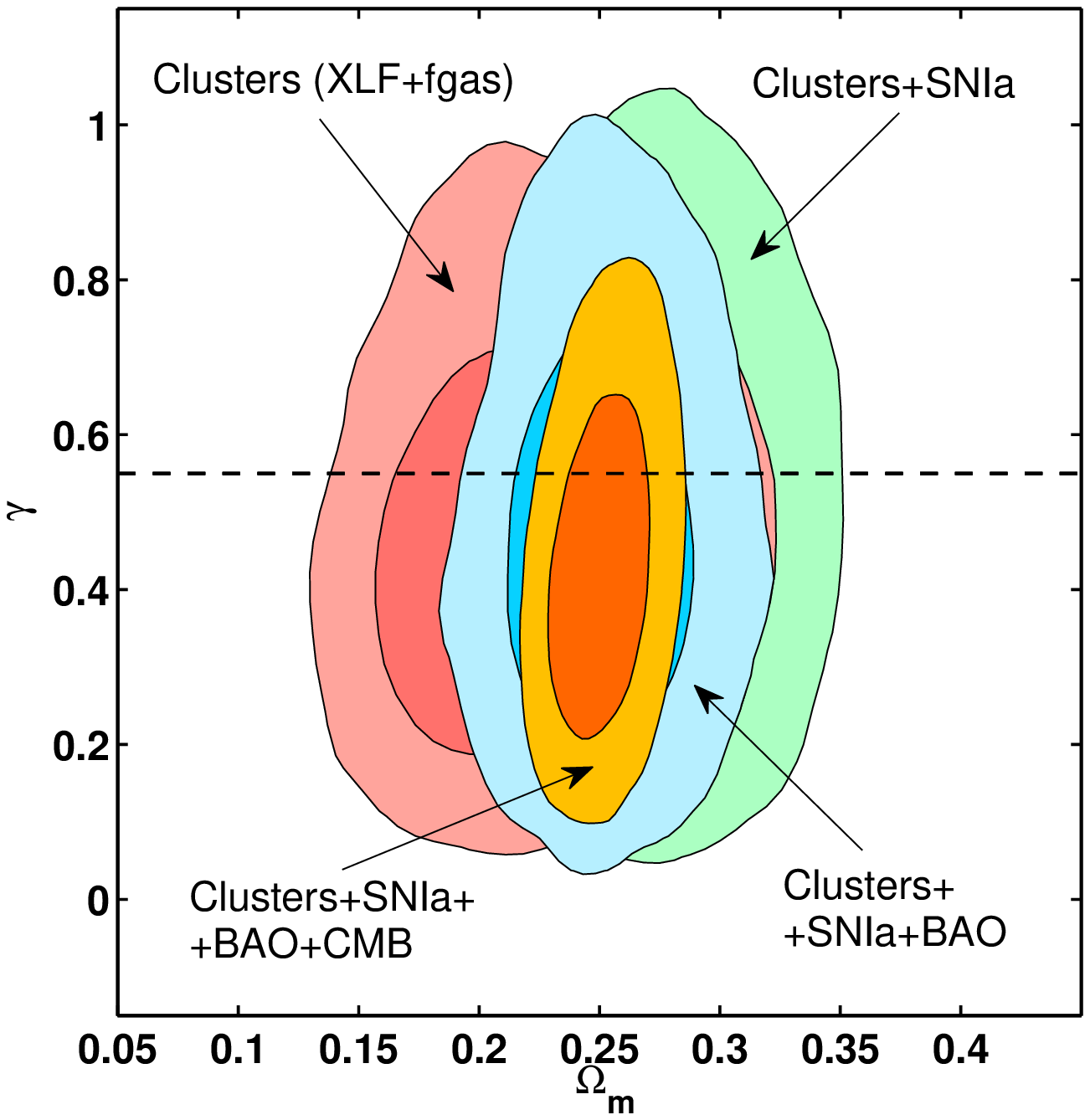}
\hspace{0.6cm}
\includegraphics[width=3.2in]{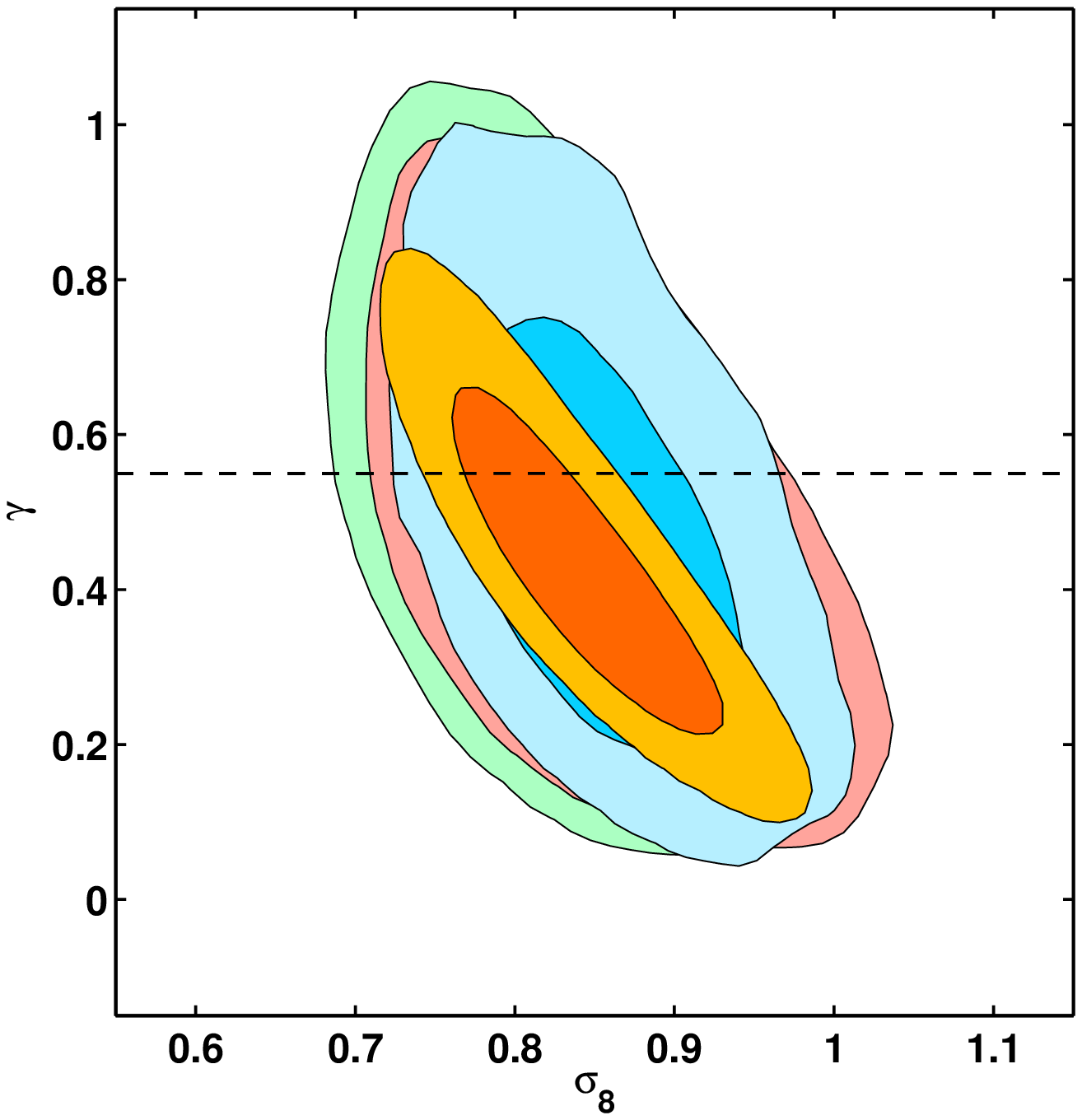}
\caption{68.3 and 95.4 per cent confidence contours in the
  $\Omega_{\rm m},\gamma$ (left panel) and $\sigma_8, \gamma$ (right
  panel) planes for the flat $\Lambda$CDM background expansion with
  self-similar evolution and constant scatter in the scaling
  relations. Results are shown for the following combinations of data
  sets: XLF+$f_{\rm gas}$ (red contours), XLF+$f_{\rm gas}$+SNIa
  (green contours), XLF+$f_{\rm gas}$+SNIa+BAO (blue contours), and
  XLF+$f_{\rm gas}$+SNIa+BAO+CMB (gold, smallest contours). This
  figure does not include the additional constraints on $\gamma$ from
  the ISW effect. The horizontal, dashed lines mark $\gamma=0.55$
  (GR).}
\label{fig:datasets}
\end{center}
\end{figure*}
 
Figure~\ref{fig:scal} shows the weak correlations between $\gamma$ and
the additional evolution parameters, $\beta_2^{\ell m}$ (left panel)
and $\sigma_{\ell m}'$ (right panel). The blue (larger) contours in
both panels are the constraints obtained with both parameters free;
the green (smaller) contours show the results for models in which only
a single additional parameter (either $\beta_2^{\ell m}$ or
$\sigma_{\ell m}'$) is allowed to vary (fixing the other to 0; see
also Table~\ref{table:params}).

In \scpaper{}, we show that, for GR models and a flat $\Lambda$CDM
background, current X-ray data at $z<0.5$ (survey+follow-up) do not
require additional evolution of the scaling relations beyond the
self-similar, constant scatter model (as implemented using the
parameters $\beta_2^{\ell m}$, $\beta^{tm}_2$, $\sigma_{\ell m}'$ and
$\sigma'_{tm}$ ; see Section~\ref{sec:scal}). Using the Deviance
Information Criterion (DIC) of \cite{Spiegelhalter:02} (see definition
in \scpaper{}) we have verified that the minimal model
($\beta^{lm}_2=0$, $\sigma_{\ell m}'=0$) remains an adequate
description of the data, when $\gamma$ is included as a parameter in
the analysis.

\begin{table*}
\begin{center}
  \caption{Marginalized 68.3 per cent confidence constraints for
    various subsets of the data, for the flat $\Lambda$CDM ($w=-1$)
    and flat $w$CDM ($w$ constant) background expansion models. Note:
    $^{a}$these results do not include the constraints on $\gamma$
    from the ISW effect.}
\label{table:params}
\begin{tabular}{ c c c c c c c }
   Data                             & $\Omega_{\rm m}$          & $\sigma_8$             & $\beta_2^{\ell m}$         & $\sigma_{\ell m}'$          & $w$                     & $\gamma$ \\
\hline                                                                                                                                                                                                           
\noalign{\vskip 4pt}                                                                                                                                                                                             
XLF+$f_{\rm gas}$                   & $0.214^{+0.036}_{-0.041}$ & $0.85^{+0.07}_{-0.06}$ & 0                     & 0                      & -1                      & $0.42^{+0.20}_{-0.16}$ \\
\noalign{\vskip 4pt}                                                                                                                                                                                             
XLF+$f_{\rm gas}$+SNIa              & $0.260^{+0.040}_{-0.025}$ & $0.83^{+0.06}_{-0.07}$ & 0                     & 0                      & -1                      & $0.43^{+0.25}_{-0.15}$ \\
\noalign{\vskip 4pt} 
XLF+$f_{\rm gas}$+SNIa+BAO          & $0.247^{+0.028}_{-0.024}$ & $0.85^{+0.05}_{-0.06}$ & 0                     & 0                      & -1                      & $0.41^{+0.24}_{-0.15}$ \\
\noalign{\vskip 4pt} 

XLF+$f_{\rm gas}$+SNIa+CMB     & $0.266^{+0.018}_{-0.021}$ & $0.85^{+0.04}_{-0.06}$ & 0                     & 0                      & -1                      & $0.40^{+0.15}_{-0.12}$ \\
\noalign{\vskip 4pt} 

XLF+$f_{\rm gas}$+SNIa+BAO+CMB$^{a}$  & $0.248^{+0.017}_{-0.012}$ & $0.84^{+0.06}_{-0.06}$ & 0                     & 0                      & -1                      & $0.42^{+0.15}_{-0.15}$ \\
\noalign{\vskip 4pt} 

XLF+$f_{\rm gas}$+SNIa+BAO+CMB & $0.249^{+0.015}_{-0.013}$ & $0.84^{+0.05}_{-0.05}$ & 0                     & 0                      & -1                      & $0.40^{+0.13}_{-0.12}$ \\
\noalign{\vskip 4pt} 

$"$  & $0.252^{+0.015}_{-0.014}$ & $0.82^{+0.05}_{-0.05}$ & $-0.36^{+0.44}_{-0.51}$ & 0                      & -1                      & $0.44^{+0.17}_{-0.13}$ \\
\noalign{\vskip 4pt} 

$"$  & $0.249^{+0.018}_{-0.010}$ & $0.82^{+0.06}_{-0.04}$ & 0                     & $-0.52^{+0.35}_{-0.33}$ & -1                      & $0.45^{+0.14}_{-0.13}$ \\
\noalign{\vskip 4pt} 

$"$  & $0.249^{+0.017}_{-0.011}$ & $0.84^{+0.05}_{-0.06}$ & $0.33^{+0.72}_{-0.76}$ & $-0.82^{+0.57}_{-0.37}$ & -1                      & $0.40^{+0.17}_{-0.13}$ \\
\noalign{\vskip 4pt} 

$"$  & $0.251^{+0.016}_{-0.014}$ & $0.83^{+0.06}_{-0.04}$ & 0                     & 0                      & $-0.98^{+0.07}_{-0.07}$ & $0.39^{+0.14}_{-0.13}$ \\

\end{tabular}
\end{center}
\end{table*}

\subsection{The impacts of the different data sets}
\label{sec:datasets}

The green, dotted-dashed line in Figure~\ref{fig:comparison} shows the
marginalized constraints on $\gamma$ from the XLF data
alone\footnote{Note that the XLF data include $f_{\rm gas}$
  measurements for the six lowest redshift clusters in the sample of
  \cite{Allen:08}, which serve to normalize the $M_{\rm gas}$ mass
  proxy (see details in Paper I). For the XLF data alone, we obtain
  the same constraints on $\gamma$ and $\sigma_{8}$ than those from
  the XLF+$f_{\rm gas}$ (full sample) data (see Table 1), and only
  $\sim 10$ weaker constraints on $\Omega_{\rm m}$.}. The red, dashed
line shows the constraints from the combination of SNIa, $f_{\rm
  gas}$, BAO, and CMB data, excluding the XLF. It is readily apparent
that the constraints on $\gamma$ from the combination of all other
data sets, excluding the XLF, are significantly weaker than those from
the XLF data alone. Without the XLF data, the constraints on $\gamma$
are primarly provided by the ISW effect, and we obtain
$\gamma=0.38^{+0.23}_{-0.69}$. Combining the XLF and other data, we
obtain the blue, solid curve and constraints on $\gamma$ more than
three times tighter than without the XLF (see
Table~\ref{table:params}).

Although the XLF data dominate the constraints on $\gamma$, the other
data ($f_{\rm gas}$, SNIa, BAO, and CMB) provide complementary
information on other cosmological parameters. These additional constraints
break degeneracies with $\gamma$ and help to exploit fully the ability
of the XLF data to constrain $\gamma$. To demonstrate the impacts of
these data sets, Figure~\ref{fig:datasets} shows the constraints in
the $\Omega_{\rm m}, \gamma$ (left panel) and $\sigma_8, \gamma$
(right panel) planes for various subsets of data. The red contours
show the constraints from the XLF+$f_{\rm gas}$ data (i.e. only
cluster data); the green contours show the results from adding SNIa
data (i.e. XLF+$f_{\rm gas}$+SNIa); the blue contours from adding BAO
to the rest; and the gold (smallest) contours from adding CMB
data\footnote{Since, for this figure, we are primarily interested in
  the ability of the CMB data to improve our knowledge of $\gamma$ by
  constraining other cosmological parameters, we do not include the
  additional constraining power on $\gamma$ from the ISW effect
  (i.e. we assume that for the large scales relevant for the ISW, GR
  is recovered). This has only a small impact on the results (see
  \citetalias{Rapetti:09} and Table~\ref{table:params}).}.

The left panel of Figure~\ref{fig:datasets} demonstrates again the
absence of any strong correlation between $\Omega_{\rm m}$ and
$\gamma$ (see also Section~\ref{sec:growth}). These results (see also
Table~\ref{table:params}) show that simply improving our knowledge of
$\Omega_{\rm m}$, and therefore the expansion history (by e.g. adding
BAO to XLF+$f_{\rm gas}$+SNIa data), will not necessarily improve the
constraints on $\gamma$.

The right panel of Figure~\ref{fig:datasets} shows the constraints in
the $\sigma_8, \gamma$ plane. We see that the combination of the CMB
data with the other data sets considerably strengthens the correlation
between these parameters. The reason for this is that the inclusion of
the CMB data significantly reduces the range of allowed values not
only for the expansion history but also for other relevant
cosmological parameters, placing tight constraints on, e.g.,
$\Omega_{\rm b}h^2$, $\Omega_{\rm c}h^2$, $n_{\rm s}$, $\tau$, $H_0$,
and $A_{\rm s}$. These constraints lead to a tight constraint on
$\sigma_{8}(z)$ at early times ($z\sim1100$), where GR is assumed
(i.e., $z>z_{\rm t}$).

\section{Conclusions}
\label{sec:conclusions}

We have used measurements of the evolution of massive galaxy clusters
from wide-area X-ray cluster surveys (spanning the redshift range
$z<0.5$), and simultaneous measurements of the observable-mass scaling
relations, to test the consistency of the observed growth rate of
clusters with that predicted for GR. We allow departures from GR as
described by the parameterization $\Omega_{\rm m}(z)^{\gamma}$, for
which $\gamma\sim0.55$ corresponds to GR. We find that the current
data are consistent with GR and give tight constraints on departures
from it. Assuming constant scatter and purely self-similar evolution
of the cluster observable-mass scaling relations, and a flat
$\Lambda$CDM background evolution model, the combination of XLF,
$f_{\rm gas}$, SNIa, BAO, and CMB data gives
$\gamma(\sigma_{8}/0.8)^{6.8}=0.55^{+0.13}_{-0.10}$. Allowing $w$ to
vary, we simultaneously demonstrate consistency with the expansion
history of $\Lambda$CDM (i.e. $w=-1$).

Our analysis employs the method described in \depaper{}, which
self-consistently models X-ray cluster survey data and follow-up
observations to simultaneously constrain both cosmological and scaling
relation parameters. This allows us to constrain $\gamma$ even when
allowing for departures from self-similar evolution and/or constant
scatter in the luminosity-mass relation. We caution that the adoption
of such a self-consistent approach to model the survey+follow-up data
is important, else spuriously tight, and potentially biased,
constraints may be obtained.

Our measurements of the growth of the most X-ray luminous, most
massive galaxy clusters within $z<0.5$ constrain the growth of cosmic
structure during the epoch of cosmic acceleration. It is primarily
this redshift dependence that provides our constraints on $\gamma$,
although we have also utilized the small contribution from the ISW
effect, which probes larger scales at $z<2$. The inclusion of CMB and
other data tightens significantly the constraints on the background
expansion model and other cosmological parameters leading to a clear
correlation between $\gamma$ and $\sigma_8$. This highlights the
potential for further improvements in $\gamma$ with the incorporation
of precise, independent measurements of $\sigma_8$.

\section*{Acknowledgments}

We thank A. Lewis, M. Amin, R. Blandford, and R. Wagoner for useful
discussions. We thank G.~Morris for technical support. The
computational analysis was carried out using the KIPAC XOC and Orange
computer clusters at SLAC. We acknowledge support from the National
Aeronautics and Space Administration (NASA) through LTSA grant
NAG5-8253, and through Chandra Award Numbers DD5-6031X, GO2-3168X,
GO2-3157X, GO3-4164X, GO3-4157X, GO5-6133, GO7-8125X\ and GO8-9118X
issued by the Chandra X-ray Observatory Center, which is operated by
the Smithsonian Astrophysical Observatory for and on behalf of the
NASA under contract NAS8-03060. This work was supported in part by the
U.S. Department of Energy under contract number DE-AC02-76SF00515. AM
was supported by a Stanford Graduate Fellowship and an appointment to
the NASA Postdoctoral Program, administered by Oak Ridge Associated
Universities through a contract with NASA.

\bibliographystyle{mnras}
\bibliography{biblist_testGR}

\label{lastpage}
\end{document}